\documentclass[preprint,12pt]{elsarticle}

\usepackage{amsmath,amssymb}    % basic symbol
\usepackage{graphicx}                           % Contains all the image-handling routines you need to include images
\usepackage{color}
\usepackage{dcolumn}
\usepackage{color}
\usepackage[usenames,dvipsnames]{xcolor}
\usepackage{url}

\journal{Physica A}

\begin{document}

\begin{frontmatter}

\title{The topology of card transaction money flows}

\author[MZ,MZ2]{Massimiliano Zanin}
\address[MZ]{Innaxis Foundation \& Research Institute,  Jos\'e Ortega y Gasset 20, 28006 Madrid, Spain}
\address[MZ2]{Departamento de Engenharia Electrot\'ecnica, Faculdade de Ci\^encias e Tecnologia, Universidade Nova de Lisboa, 2829-516 Caparica, Portugal}

\author[URJC]{David Papo}
\author[URJC,CTB]{Miguel Romance}
\author[URJC,CTB]{Regino Criado}
\address[URJC]{Department of Applied Mathematics, Universidad
  Rey Juan Carlos,  28933 M\'ostoles, Madrid, Spain}
\address[CTB]{Center for Biomedical Technology, Universidad
  Polit\'ecnica de Madrid,  28223 Pozuelo de Alarc\'on, Madrid, Spain}

\author[BBVA]{Santiago Moral}
\address[BBVA]{Cyber Security \& Digital Trust, BBVA Group,  28050 Madrid, Spain}

\begin{abstract}
Money flow models are essential tools to understand different economical phenomena, like saving propensities and wealth distributions. In spite of their importance, most of them are based on synthetic transaction networks with simple topologies, {\it e.g.} random or scale-free ones, as the characterisation of real networks is made difficult by the confidentiality and sensitivity of money transaction data. Here we present an analysis of the topology created by real credit card transactions from one of the biggest world banks, and show how different distributions, {\it e.g.} number of transactions per card or amount, have nontrivial characteristics. We further describe a stochastic model to create transactions data sets, feeding from the obtained distributions, which will allow researchers to create more realistic money flow models.
\end{abstract}


\end{frontmatter}

\section{Introduction}
\label{sec:intro}

One of the many relevant open problems in economics, and more recently in econophysics, is the understanding of the basic mechanisms driving income or wealth distribution in various economic communities, for instance in different countries.
While it has been shown that the associated probability distribution for the whole population roughly follows a power-law, also known as Pareto distribution \cite{reed2001pareto}, and that the low-income population is better described by a Gibbs distribution \cite{dragulescu2002statistical}, less is known about the mechanisms generating these distributions.

The models proposed in the literature \cite{andresen1998macroeconomy}, typically include the money flow between agents, and their saving propensity. Most of the studies consider random interactions between agents \cite{dragulescu2002statistical,wang2003circulation}, or even mean-field approximations \cite{beyeler2006congestion,tamura2012estimation}, in which all agents interact with all others, two clearly unrealistic assumptions that do not reflect the reality of economic transactions. A few studies allow agents to interact above more complex topologies \cite{souma2001small,souma2003wealth,hu2006unified,hu2007simulating}. However,  the synthetic network models ({\it e.g.} scale-free networks \cite{barabasi1999emergence}) are not based on real observations.
The information on everyday monetary transactions is often considered sensitive. The resulting scarcity of available data  represents a severe limitation and explains why very few studies have tackled this issue, mainly resorting to the possibilities offered by new on-line technologies, {\it e.g.} BitCoins \cite{kondor2014do}.

The modelling of money flows bears some similarities with that of modelling the information or opinion spreading in a social network. In this field of research too the initial models only considered simple topologies \cite{nowak1990from, Sobkowicz2009Modellin}, but it was soon realised that the structures created by interactions between people \cite{wu2004information} and the time-varying organisation of these interactions \cite{iribarren2009impact, miritello2011dynamical, miritello2013temporal} both have a dramatic impact on how faithfully the models translates the underlying spreading process.

Thus, in analogy with the approach followed in social sciences, one may ask whether and to what degree the transaction topologies are important for creating realistic money flow models.
In this contribution, we make one step in this direction by describing and analysing a unique large-scale data set, including all credit and debit card transactions realised in Spain between 2011 and 2012 within the network of one of the biggest Spanish banks. This includes information for both transactions identified as normal and for those identified as fraudulent.

We assume that money flows can be represented as a two-layer network \cite{boccaletti2014structure}, with users on one layer, and stores (or more generally, any economic activity) on the other. Card transactions then represent flows going from the first to the second layer. To this, other types of flows should be added: inter-users ({\it e.g.} transfer of money among members of a family); inter-stores ({\it i.e.} commercial activities); and from stores to users ({\it i.e.} salaries). Clearly, the analysis of card transactions alone allows a limited representation of the whole system; furthermore, it does not account for transactions executed in cash.
While this analysis presents some important limitations, it has to be noted that it is, to the best of our knowledge, the first of its kind; while this reduced view is in line with previous studies \cite{kondor2014do}, it represents at the same time a picture of the real economy, and not just a virtual one. Additionally, the possibility of analysing the dynamics of illegal transactions enables a first attempt to characterise the ``black economy'', something which has seldom been done in the past.

The content of the work is organised as follows. Section \ref{sec:transactions} describes the main characteristics of the legal transactions, and thus characterises the network topology of links created between the user and the store layers. Section \ref{sec:illegal} presents some results related to illegal transactions. Finally, some conclusions are drawn in Section \ref{sec:discussion}.

\section{Transactions analysis}
\label{sec:transactions}

As previously introduced, the studied data set accounts for all credit and debit card transactions of clients of the Spanish bank BBVA, from January 2011 to December 2012. Each month, an average of $15$ million operations were realised by $7$ million cards, for a total of $250$ GB of information. Available fields included a time stamp of the operation, the quantity (both in Euro and in the original currency, if different), and the origin (the card) and destination (the store) of the operation; the two latter fields were anonymised, so that the exact card number and the name of the store could not be recovered. Additionally, information on the outcome of the operation was included, {\it i.e.} whether the operation was detected as illegal or denied for any other reason. In this section, only legal and accepted operations are examined.

\begin{figure}
\begin{center}
\includegraphics[width=0.9\textwidth]{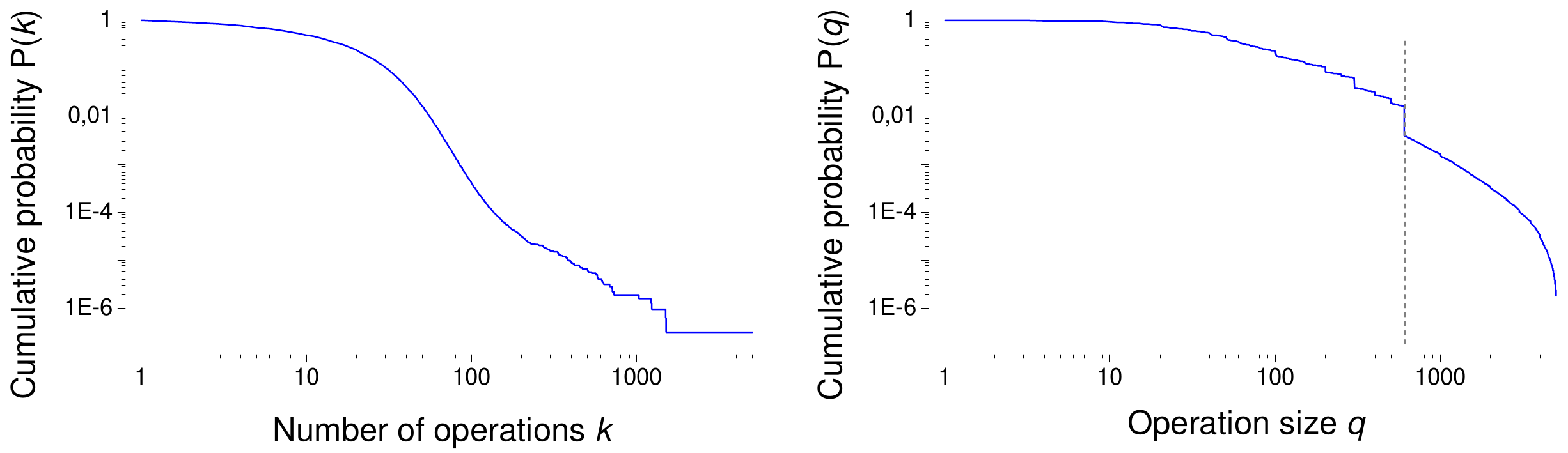}
\end{center}
\caption{\label{fig01} Analysis of credit card transactions. (Left) Cumulative probability distribution of the number of operations realised by different cards. (Right) Cumulative probability distribution of the amount (in Euro) of each transaction.}
\end{figure}

Fig. \ref{fig01} Left and Right respectively represent the cumulative probability distribution (CPD) of the number of operations per card, and the operation size (in Euro). The CPD, with respect to a given parameter value $v$, accounts for the probability of finding an operation with at least value $v$, {\it i.e.} $P(v) = p( v_i > v)$ with $i$ running all over operations. In Fig. \ref{fig01} Left, $k_i$ is thus the number of operations performed by card $i$ in one month; on the other hand, in Fig. \ref{fig01} Right, $q_i$ is the quantity of money involved in operation $i$. While both graphs suggest a scale-free distribution, with long tails, this approximation is far from being precise. Specifically, the number of operations per card seems to display three regions of different slope: between $1$ and $20$, between $20$ and $200$, and above $200$ operations per months, suggesting that different types of users are included. This is further confirmed by Fig. \ref{fig01} Right, in which there is a marked jump at $600$ Euro; this is due to the fact that this quantity is, by default, the maximum that can be withdrawn with a debit card in Spain.

\begin{figure}
\begin{center}
\includegraphics[width=0.9\textwidth]{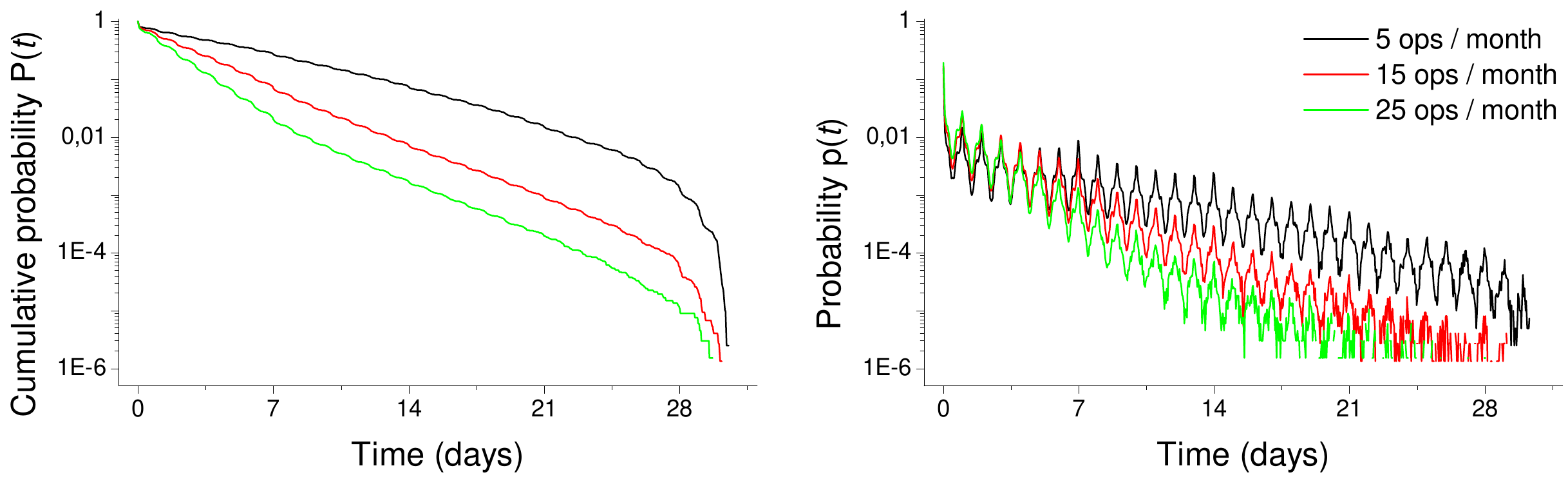}
\end{center}
\begin{center}
\includegraphics[width=0.4\textwidth]{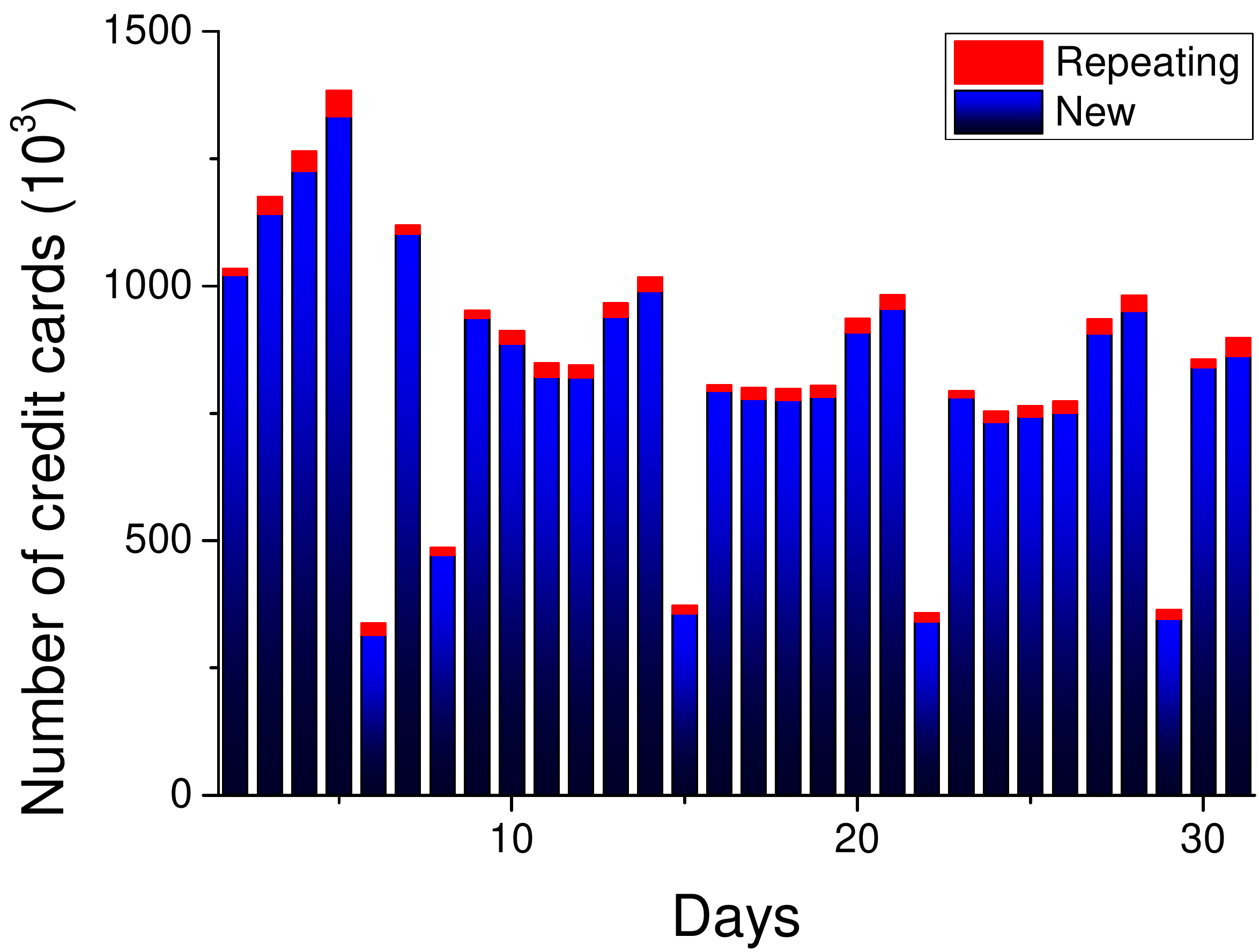}
\end{center}
\caption{\label{fig02} Analysis of user behaviours. Top Left and Right panels respectively depict the cumulative and normal probability distribution of the time passed between two consecutive operations. Black (top), red (middle) and green (bottom) curves respectively represent the average for users with 5, 15 and 25 operations per month. The Bottom panel presents the number of cards operating per day, divided between new cards ({\it i.e} those that did not operate the previous day, red bars) and repeating ones (blue bars).}
\end{figure}

Beyond the description of the static, or averaged, topology, it is also of interest to evaluate how transactions change and evolve through time. In particular, it is important to examine whether decisions of using a card are related to past decisions. To shed light on this issue, Fig. \ref{fig02} Top depicts the cumulative (Left panel) and normal (Right panel) probability distribution corresponding to the time passed between two consecutive operations of the same card. Three curves are represented, corresponding to cards which have realised $5$, $15$ and $25$ operations in a single month - thus representing the spectrum of normal users.
A clear oscillatory dynamics of 24h period is present in Fig. \ref{fig02} Top Right - which is partly smoothed out in the cumulative representation. This indicates that, especially for a low number of operations, users are characterised by a periodic behaviour, in which similar operations are performed at the same time of the day - {\it e.g.} one may tend to visit the gas station at the same hour in the morning, or pay a parking always after work.
Fig. \ref{fig02} Bottom further depicts how cards reappear in consecutive days, by showing the number of new cards ({\it i.e} those that did not operate the previous day) and of repeating ones. It can be seen that most of them are not repeating in consecutive days, indicating that most of the users do not use the same card within two days.

\begin{figure}
\begin{center}
\includegraphics[width=0.7\textwidth]{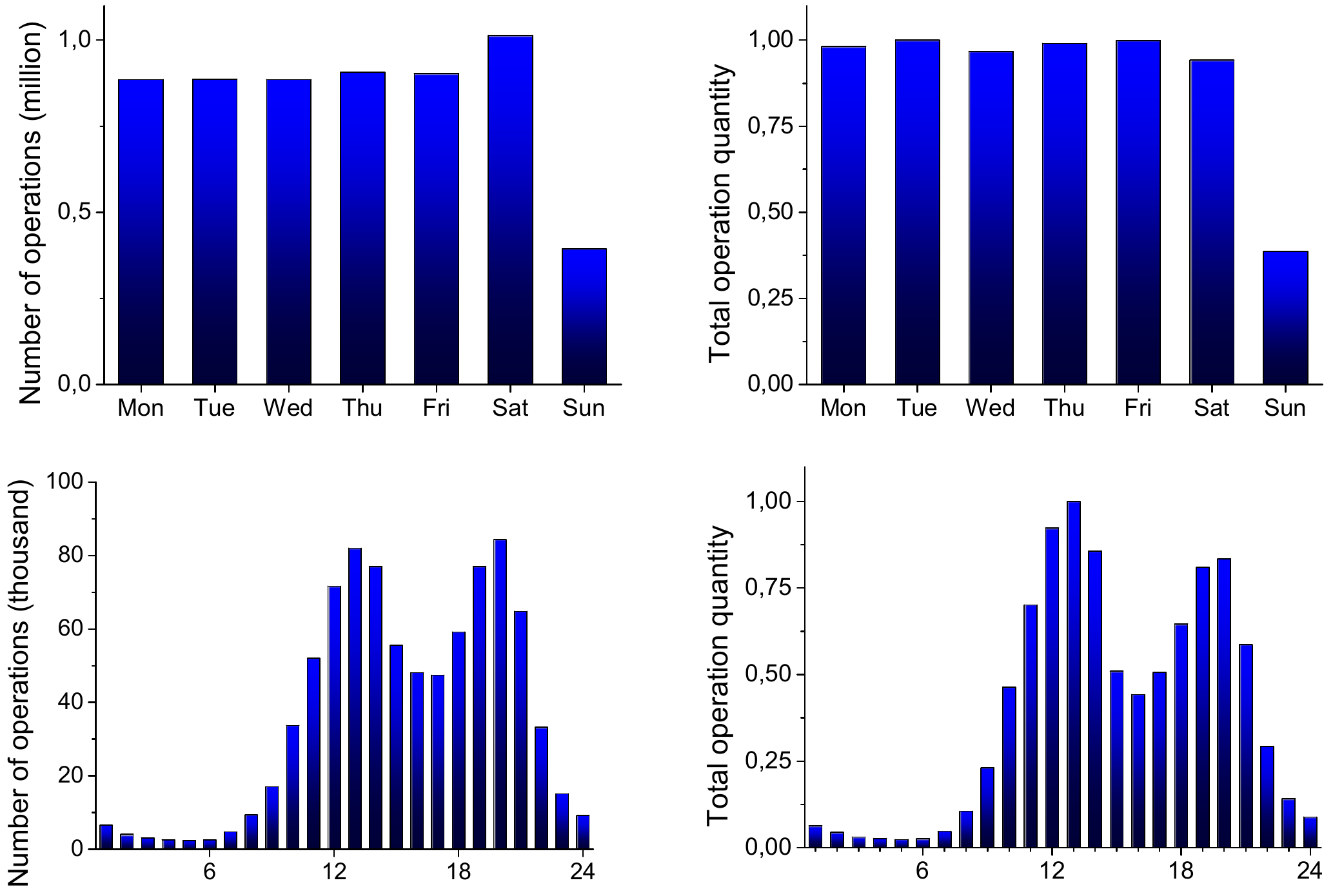}
\end{center}
\caption{\label{fig03} Distribution of hourly and daily activities. Top and bottom graphs respectively represent the distribution of operations during the seven days of the week, and the 24 hours of a day. Left panels correspond to the number of transactions, right ones to the normalised amount of operations.}
\end{figure}

Finally, Fig. \ref{fig03} shows how operations depend on the day of the week (Top panels) and on the hour of the day (Bottom panels). As may be expected, operations are reduced during Sunday, but not during Saturday - as people use their credit cards in their leisure activities. There is also an important reduction of the card activity during night hours, with two marked peaks at Spanish lunch and dinner time (13:00 - 14:00 and 20:00 - 22:00).

\section{Illicit transactions}
\label{sec:illegal}

\begin{figure}
\begin{center}
\includegraphics[width=0.9\textwidth]{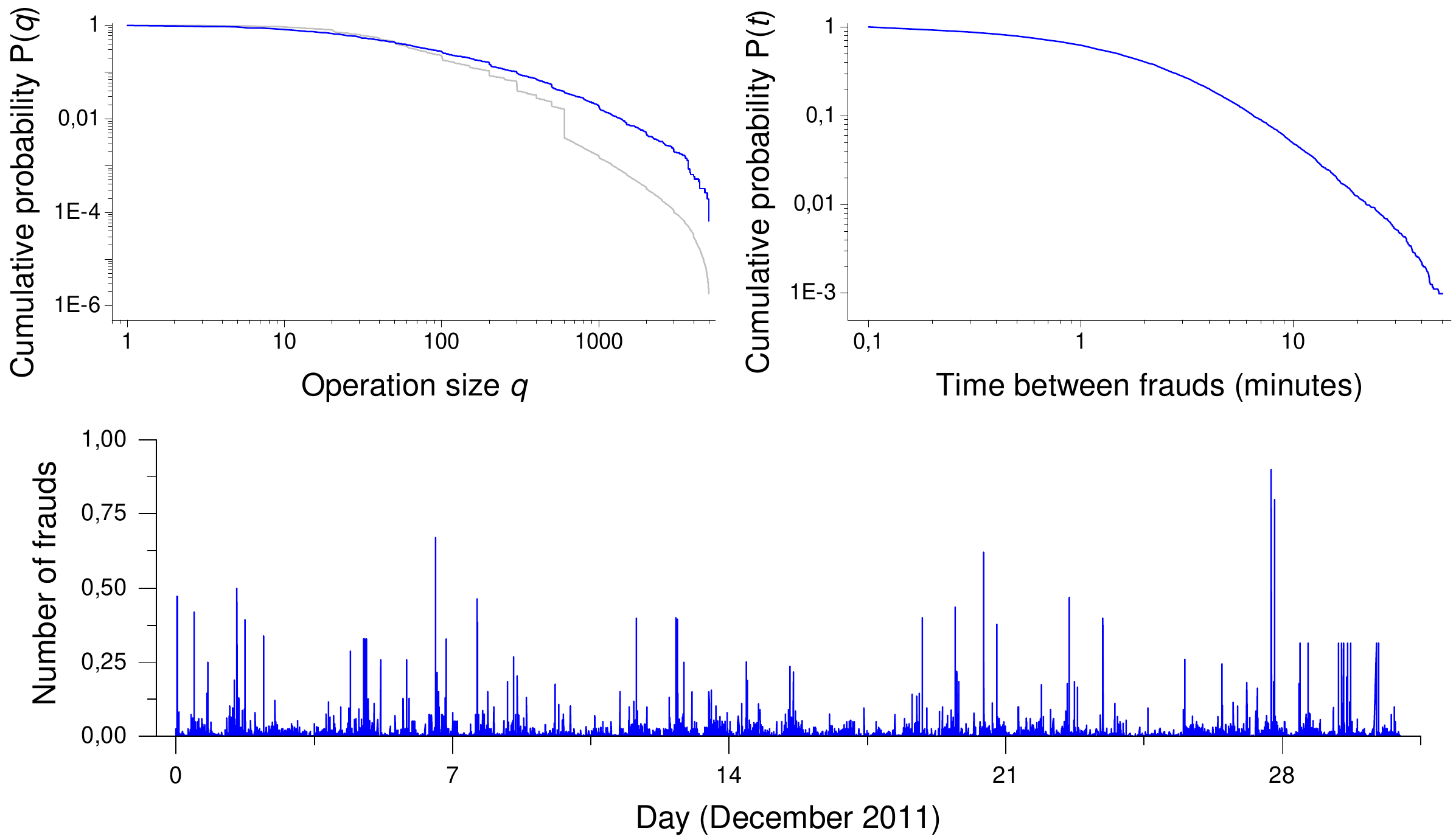}
\end{center}
\caption{\label{fig04} Analysis of illicit transactions. (Top left) Cumulative probability distribution of the amount (in Euro) of illegal transactions; the curve corresponding to legal ones is reported in grey. (Top right) Cumulative probability distribution of the time between two consecutive illegal transactions. (Bottom) Time evolution of the normalised number of transactions for December 2011.}
\end{figure}

In this section, we focus on those transactions that were marked as illicit by the bank's fraud prevention software\footnote{Due to confidentiality issues, details about this system have been omitted.}. Fig. \ref{fig04} Top Left presents the CPD of the operation size (in blue), compared to the one obtained for legal transactions (light grey). The former distribution presents a higher probability for large transaction sizes, suggesting that attackers prefer to operate large transactions - which are more profitable when successful. Additionally, Fig. \ref{fig04} Top Right depicts the CPD for the time between consecutive illegal transactions - all transactions are considered together, as the identity of the attacker is not known; and Fig. \ref{fig04} Bottom depicts the evolution of the number of illegal transactions during December 2011\footnote{Due to confidentiality reasons, the number of illegal transactions has been normalised in the range $[0, 1]$.}.

\section{A model for synthesising transactions}
\label{sec:model}

The problem of data confidentiality seldom allows researchers to test a money flow model on a real data set, {\it i.e.} describing real transactions taken place within an economical system. In order to bridge this gap, we here present a model able to synthesise a set of transactions starting from the real probability distributions previously obtained. In this section, we are going to describe its general organisation - a more detailed description of its functions can be found in \ref{sec:Annex1}, and its validation is discussed in \ref{sec:Annex2}.

The model starts by defining a set of initial credit cards and stores. Each one of them is characterised by an identification number, and by an expected size, {\it i.e.} the number of transactions that are expected to take place in a given time window. No restriction is imposed on the relation between the number of cards and stores. Due to this, the size of the former is strict, meaning that each card is expected on average to execute the assigned number of transactions; on the contrary, the size of stores is relative, and is used only in a proportional way. This flexibility allows creating synthetic data sets with different sizes, {\it i.e.} from the original system up to small test cases.

Once both sets have been initialised, the algorithm creates transactions in a sequential manner. For each day in the specified time window, the algorithm sequentially selects all cards, one by one, calculates the number of expected transactions (also taking into account the day of the week), and stores those transactions in memory. To these, three more attributes are linked: the hour of the day (according to the provided distribution), the store, and the size of the transaction. Notice how these three attributes are independent from the card executing the transactions; in other words, second order distributions are not considered. In spite of this, the model is able to recreate all the most important features of the real transaction set - see \ref{sec:Annex2} for the results of the validation process.

The synthetic model further includes two options, which allow the creation of a more rich variety of user behaviours.
The first one is designed to reduce the determinism of the model, and allows introducing a ``burst" behaviour in card dynamics. Suppose the model is simulating a credit card, expected to execute 15 operations per month. This corresponds to $0.5$ operations per day, or, in other words, on average one operation every two days. The real behaviour of transactions is nevertheless different, as represented in Fig. \ref{fig03}: even cards expected to execute few operations every month have some probability of executing those transactions in a single day, {\it i.e.} to present a ``burst" dynamics. In order to simulate this, the model can introduce a tuneable amount of variability in the number of daily transactions; if a card executes more transactions than the expected average, an inhibition is saved for the following days, in order to recover the equilibrium.

The second option allows to randomly swap cards, and introduces a higher degree of heterogeneity in the model. It consists of a random process that exchange the expected average activity of two cards with a given probability, if some constraints ({\it e.g.} similarity between cards) are fulfilled.

\section{Conclusions}
\label{sec:discussion}

Network science has had a fabulous development in the last twenty years, and is having a huge impact on a multitude of fields, from social sciences to biomedicine \cite{costa2011analyzing}. However, to the best of our knowledge, no model of credit card transactions has been proposed based on a multi-layer complex network point of view \cite{boccaletti2014structure}. In this work, we have described a stochastic model to create transaction data sets, feeding from the distributions obtained after a detailed analysis of the topology created by real credit card transactions from one of the biggest world banks between 2011 and 2012. Unlike other models proposed in the literature, which typically consider random interactions and money flows between agents, our approach assumes that money flows can be represented as a two-layer network, with users laying on one layer and stores on the other, and that card transactions represent flows going from the first to the second layer. In the future we plan to explore other approaches based on larger and more detailed multilayer network representations, {\it e.g.} by considering different layers for different types of stores, and how users can be classified in a number of ways. We hope our model will allow researchers to create more realistic money flow models.

\section{Acknowledgements}

 The authors gratefully acknowledge the Technological Risk Management Research Center (Centro para la Gesti\'on Tecnol\'ogica del Riesgo, CIGTR) sponsored by the Rey Juan Carlos University and BBVA group, and the I4S-URJC Chair on Information Security, Fraud Prevention and Technological Risk Management that encourage the base research on pain points of risk management within information systems. 
 
This work has been partly supported by the Spanish MINECO under project  MTM2014-59906-P and by the grant for the researching activity for excellence group GARECOM GI\_EXCELENCIA 30VCPIGI11.

\bibliographystyle{model1-num-names}

\appendix

\section{Synthetic transactions model: structure and usage}
\label{sec:Annex1}

The model has been developed as a set of Python libraries, which can be downloaded at \url{www.mzanin.com/MoneyFlow}. Files, and functions within, are organised as follows:

\begin{description}
	\item \emph{SynthTransactions}. Main file of the library, it contains the following functions:
		\begin{itemize}
			\item \emph{CreateTransactions}. This is the main function that creates the synthetic data set. Several parameters are accepted as input: the number of users and stores, the time window covered by the data set, {\it etc.} - see the file itself for a full list. The output is a list, each element describing a single transaction and containing five fields: the day, the identification of the user, the hour, the quantity involved in the transaction, and the identification of the store.
			\item \emph{NormalizeDistributions}. Internal function that normalises distributions.
		\end{itemize}
	\item \emph{SynthTransactions\_InputData}. This file includes the definition of the variables encoding the probabilities distributions to be used in the transaction synthesis. Values provided in the original file represent real transactions, but can be changed as needed. They include: 
		\begin{itemize}
			\item \emph{HourlyDistr}, the distribution of transactions across the 24 hours of the day.
			\item \emph{DailyDistr}, the distribution of transactions across the seven days of the week.
			\item \emph{QuantityDistr}, 50 values that describe the distribution of transaction sizes, in steps of 25 monetary units.
			\item \emph{NumOpsDistr}, 100 values describing the probability distribution of the monthly number of transactions per card.
			\item \emph{NumOpsDistrStores}, describing the probability distribution of the number of transactions of stores. 50 values are expected, in steps of 20.
		\end{itemize}
	\item \emph{SynthTransactions\_TestFunctions}. Set of functions allowing the recreation of the graphs presented in this paper. The \emph{matplotlib} Python library is used to plot the results. The obtained distributions are also returned for further processing.
	\item \emph{TestProgram}. A simple test program, showing how to setup and run the model.
\end{description}

In order to create a synthetic data set, it is only necessary to run the main \emph{CreateTransactions} function with a suitable set of parameters. For instance, the following code allows creating a data set comprising 2000 cards, 1000 stores, and transactions corresponding to 100 days:

\begin{quote}
	\begin{verbatim}
Transactions = CreateTransactions(2000, 1000, 100)
	\end{verbatim}	
\end{quote}

A sample of the result, including three transactions and as stored in \emph{Transactions}, is here reported:

\begin{quote}
	\begin{verbatim}
(0, 1, 17, 87.5, 78)
(0, 1, 17, 37.5, 544)
(0, 2, 13, 62.5, 68)
	\end{verbatim}	
\end{quote}

\section{Synthetic transactions model: validation}
\label{sec:Annex2}

\begin{figure}
\begin{center}
\includegraphics[width=0.37\textwidth]{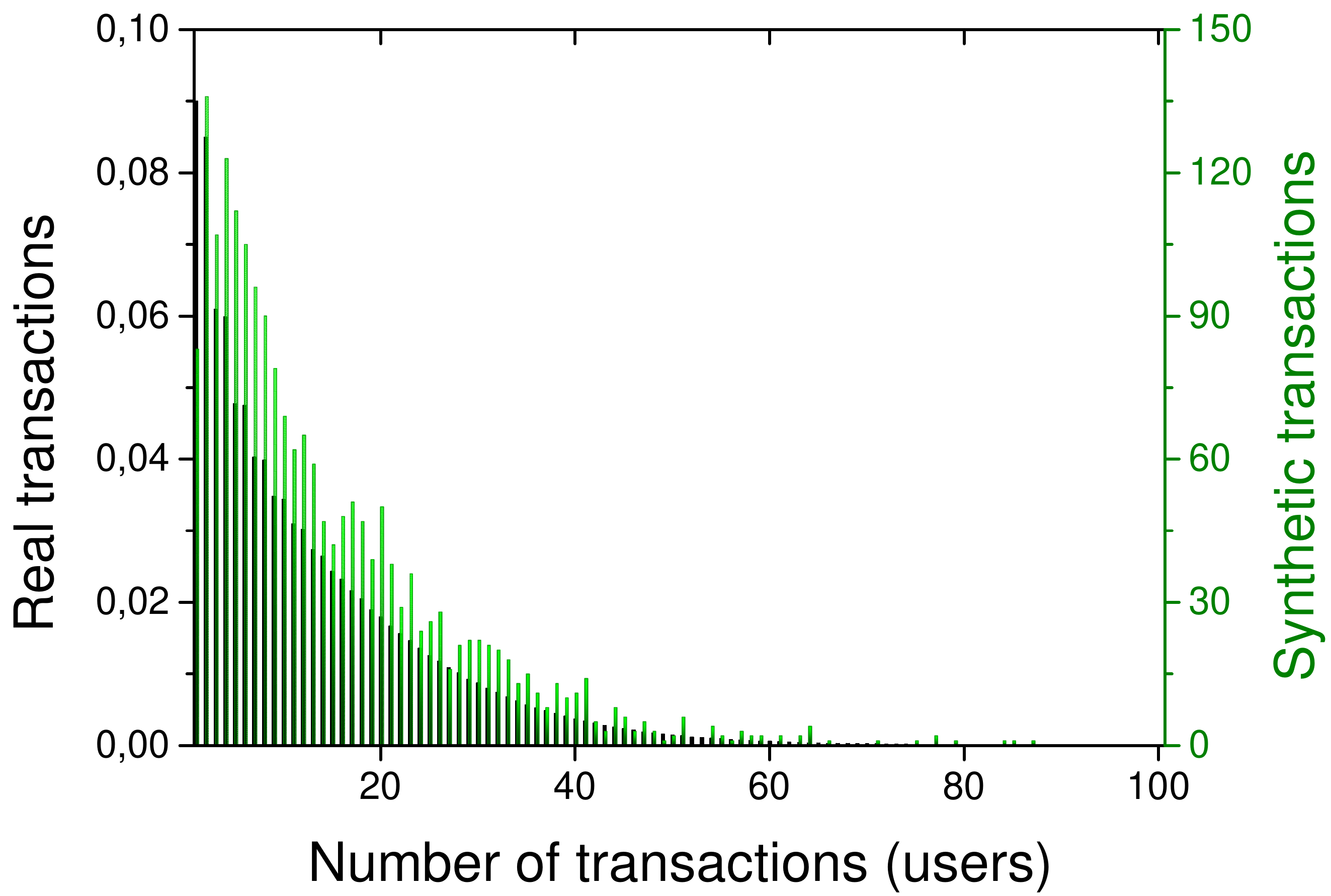}
\hspace{0.5cm}
\includegraphics[width=0.37\textwidth]{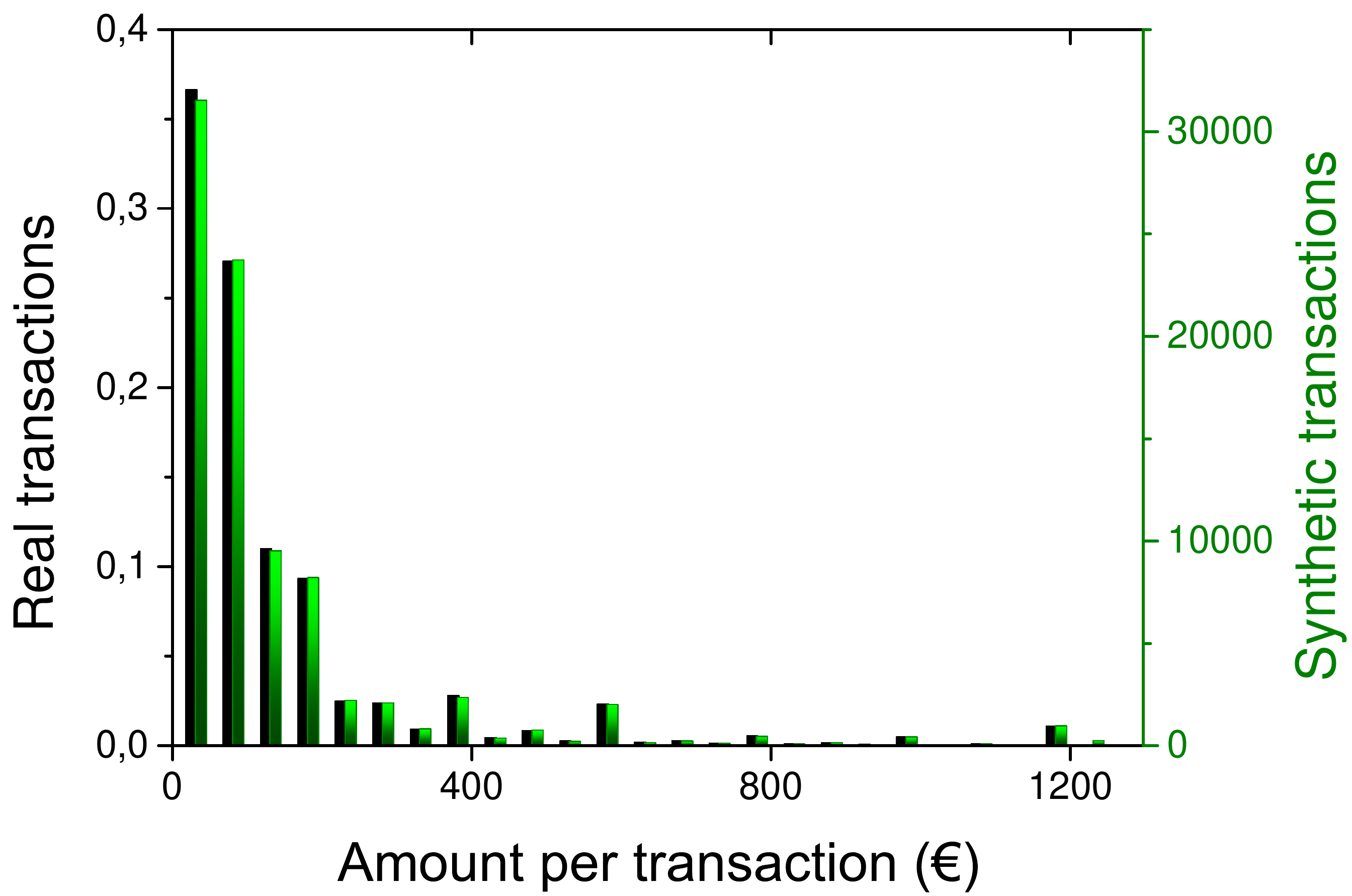}
\vspace{0.3cm}
\includegraphics[width=0.37\textwidth]{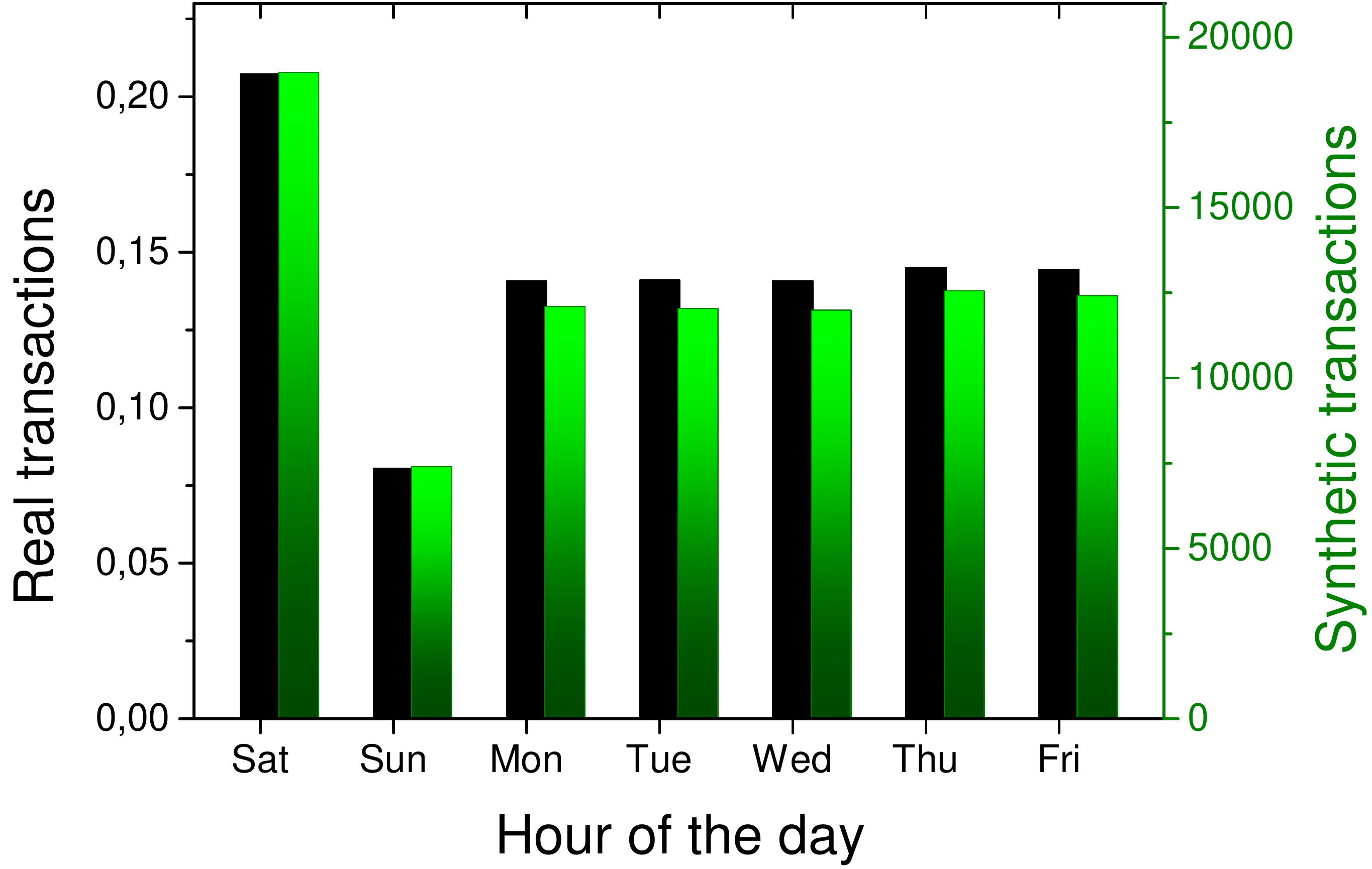}
\hspace{0.5cm}
\includegraphics[width=0.37\textwidth]{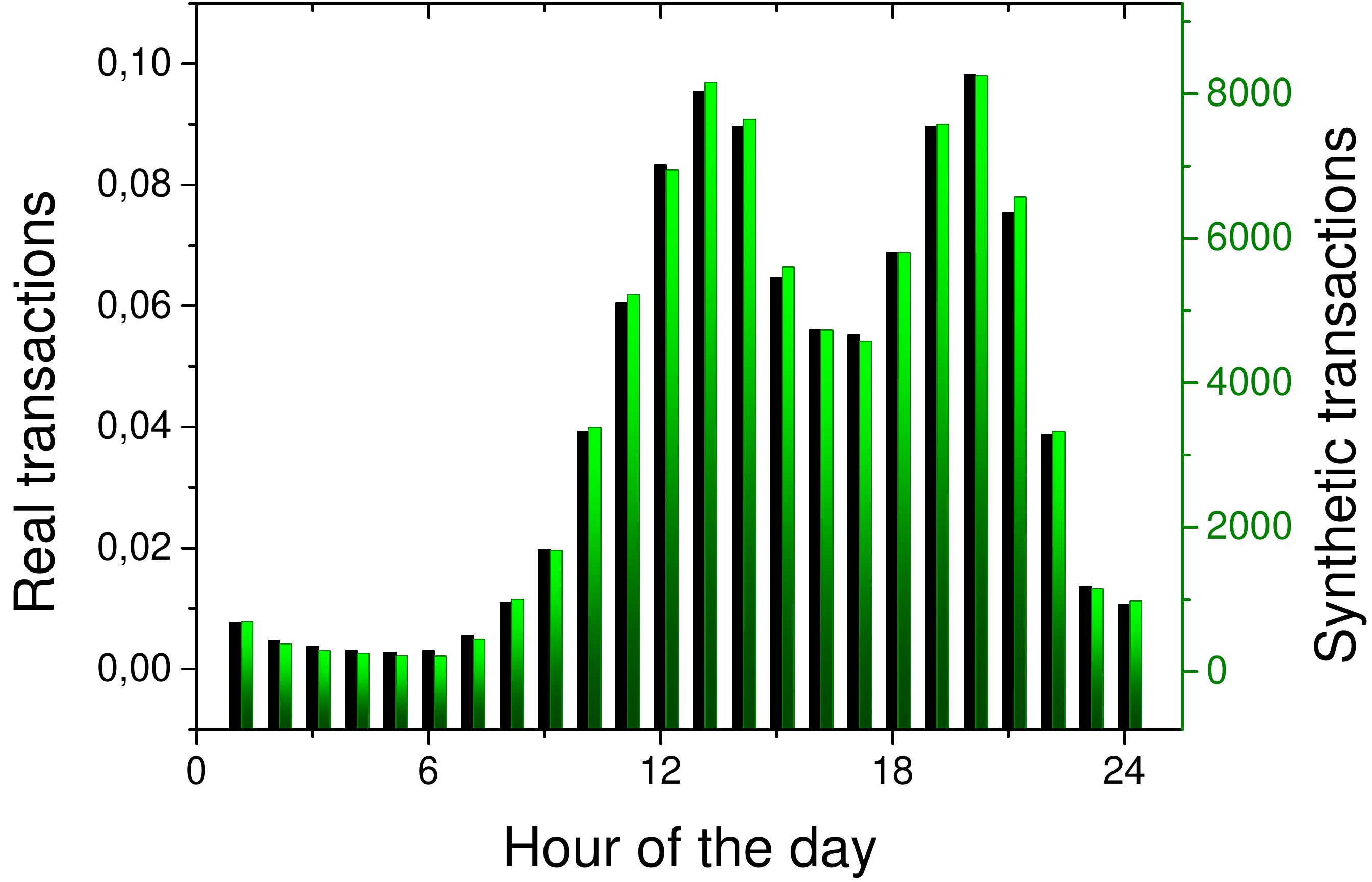}
\vspace{0.3cm}
\includegraphics[width=0.8\textwidth]{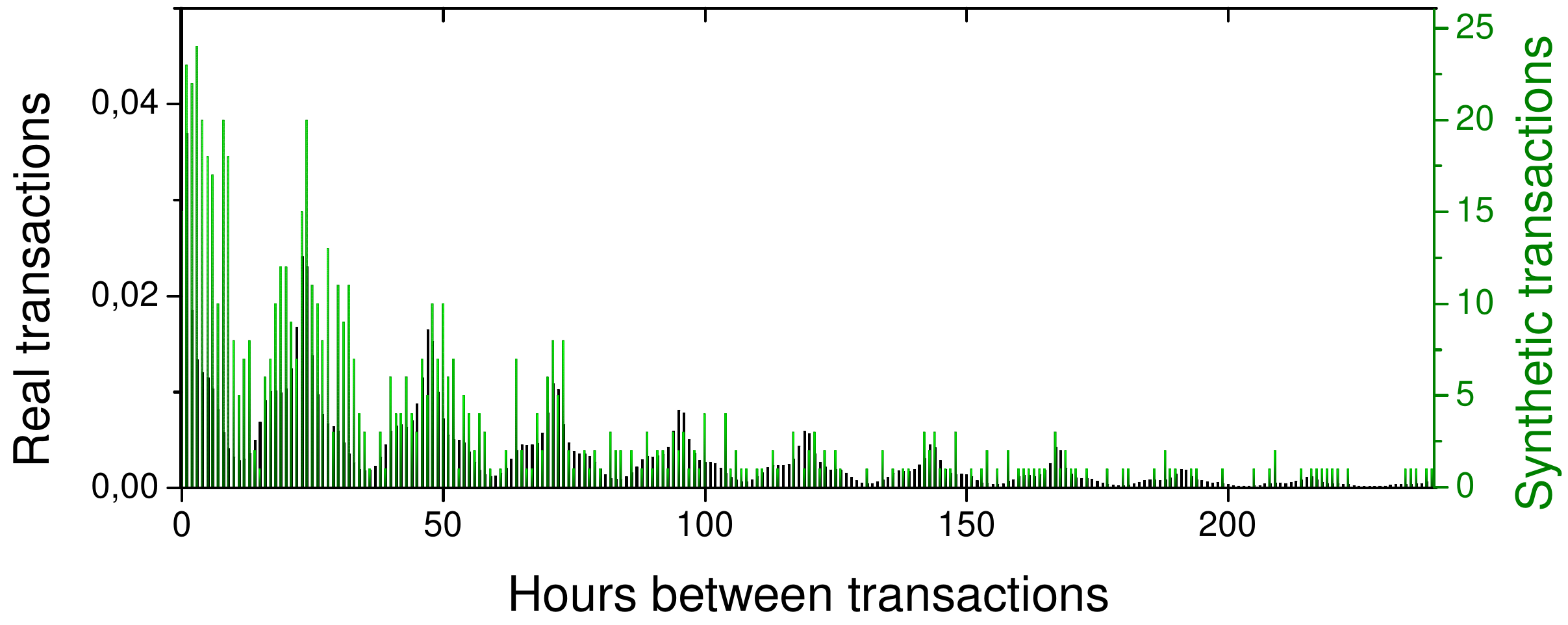}
\vspace{0.3cm}
\includegraphics[width=0.8\textwidth]{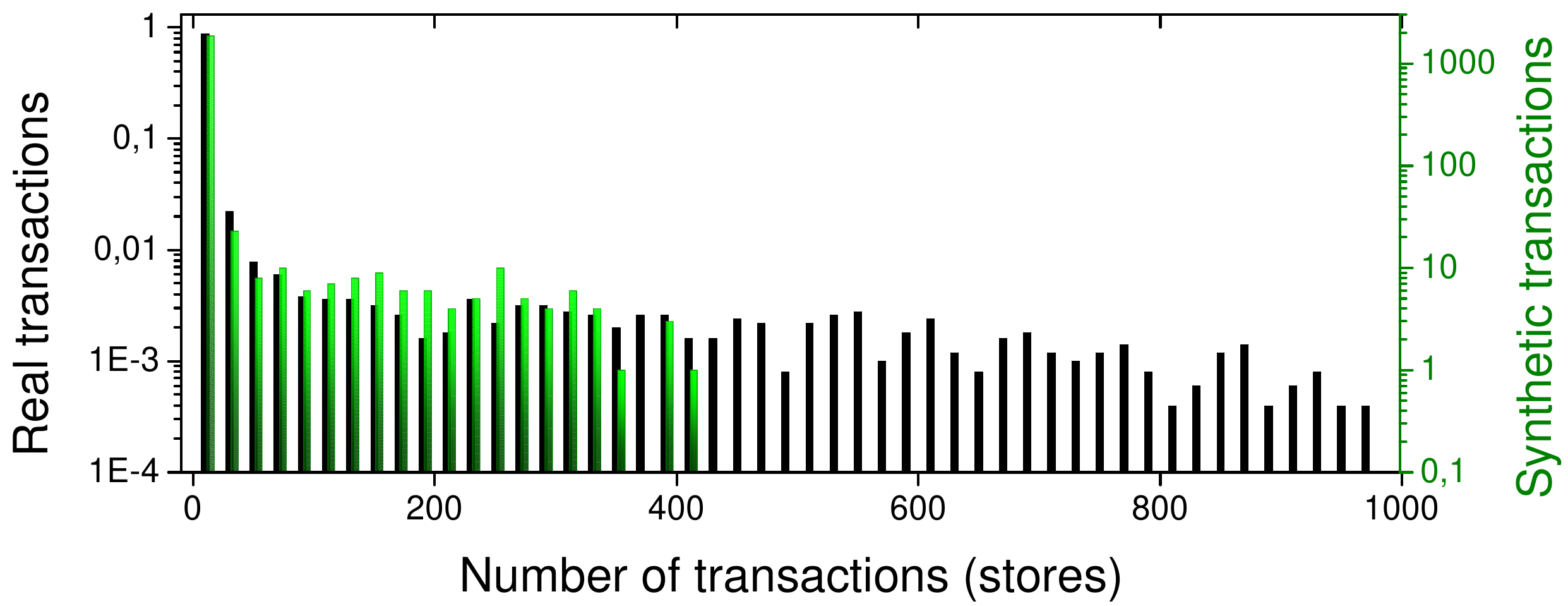}
\end{center}
\caption{\label{figAux01} Validation of the synthetic model for creating transaction sets. Black and green bars respectively represent the probability distributions obtained from real and synthetic data. From left to right, top to bottom, the six panels represent: ({\it i}) the distribution of the number of transactions per user (credit card); ({\it ii}) distribution of amount of each transaction; ({\it iii}) distribution according to the day of the week; ({\it iv}) distribution according to the hour of the day; ({\it v}) distribution between consecutive transactions of the same card; and ({\it vi}) distribution of the number of transactions per store.}
\end{figure}

The validation of the model has been executed by creating a set of synthetic transactions, and by comparing the resulting probabilities distributions with the one observed in real data. Results are presented in Fig. \ref{figAux01}, for a data set comprising $2.000$ cards, $1.000$ stores, and $100$ days, for a total of $92.500$ transactions. Black and green bars respectively correspond to the distributions obtained for real and synthetic transactions.

Both sets of distributions present remarkable similarities - notice that some of them, like the distribution of the time between subsequent transactions of a same card, are not hard-coded in the model. The differences that can be observed in the last panel of Fig. \ref{figAux01} are due to the limited number of stores, which creates a finite size effect.

\end{document}